\documentclass{article}
\usepackage[utf8]{inputenc}

\usepackage{graphicx}
\usepackage{subcaption}
\usepackage[export]{adjustbox}
\usepackage{wrapfig}
\usepackage{algpseudocode}

\usepackage{hyperref}
\usepackage{multicol} 
\providecommand{\keywords}[1]  
{
  \immediate	
  \textbf{\textit{Keywords---}} #1
}
\usepackage{authblk}

\usepackage{tabularx} 

\usepackage{array}
\newcolumntype{C}[1]{>{\centering\arraybackslash}m{#1}}

\title{Detecting Phishing sites Without Visiting them}

\author{Kalaharsha Pagadala $^{a}$ \\
        \small $^{b}$School of Computer Science and Information Sciences
(SCIS), University of Hyderabad, Hyderabad, India

} 

\date{}

\begin{document}

\maketitle

\begin{abstract}
Now-a-days, cyberattacks are increasing at an unprecedented rate. Phishing is a social engineering attack which has a massive global impact, destroying the financial and economic value of corporations, government sectors and individuals. In phishing, attackers steal users personal information such as username, passwords, debit card information and so on. In order to detect zero-hour attacks and protect end-users from these attacks, various anti-phishing techniques are developed, but the end-users have to visit the websites to know whether they are safe or not, which may lead to infecting their system. In this paper, we propose a method where end-users can detect the genuineness of the sites without visiting them. The proposed method collects legitimate and phishing URLs and extract features from them. The extracted features are given as input to six different classifiers for training and constructing the model. The classifiers used are Naive-Bayes, Logistic Regression, Random Forest,CatBoost, XGBoost and Multilayer perceptron. The method is tested by developing into an extension so that the end-users can use it when browsing. In the browser extension when the user takes the cursor over any link, a pop-up appears showing the nature of the website i.e., safe site or deceptive site and then a confirm box shows up asking the user whether they want to visit or not. The performance of the approach is tested using a dataset consisting of 2000 phishing and legitimate website URLs and the method is able to detect the sites correctly in very little time. Random-Forest is chosen for constructing the model as it gives the highest accuracy of 95\%.
\end{abstract}

\section{Introduction}
Phishing is one of the major cyberattacks [2] prevailing at a faster rate. In phishing, attackers lure the end-users by making them click the hyper-links which make them lose their personally identifiable information, banking information such as credit card details and passwords. In this attack the attackers disguise themselves as trusted entities such as service providers, employees of the organization so that end-users never doubt them. It is mainly done through emails asking to update the system, or saying that the account has been suspended, or asking to claim the prize and so on [3]. The main goal of phishing is to make end-users share their sensitive information. Websites play a pivot role in phishing attacks as they provide services like providing information, providing personal accounts to store, buy or sell the services and can be used to lure customers to steal their data easily.
\newline
\newline
According to Anti-Phishing Working Group (APWG) [1], Phishing Activity Trends Report, 4th Quarter 2020 report states that the number of phishing attacks observed by APWG and its members grew through 2020, doubling over the course of the year. 
\begin{figure}[htpb!]
  
    \centering 
\includegraphics[scale=0.3]{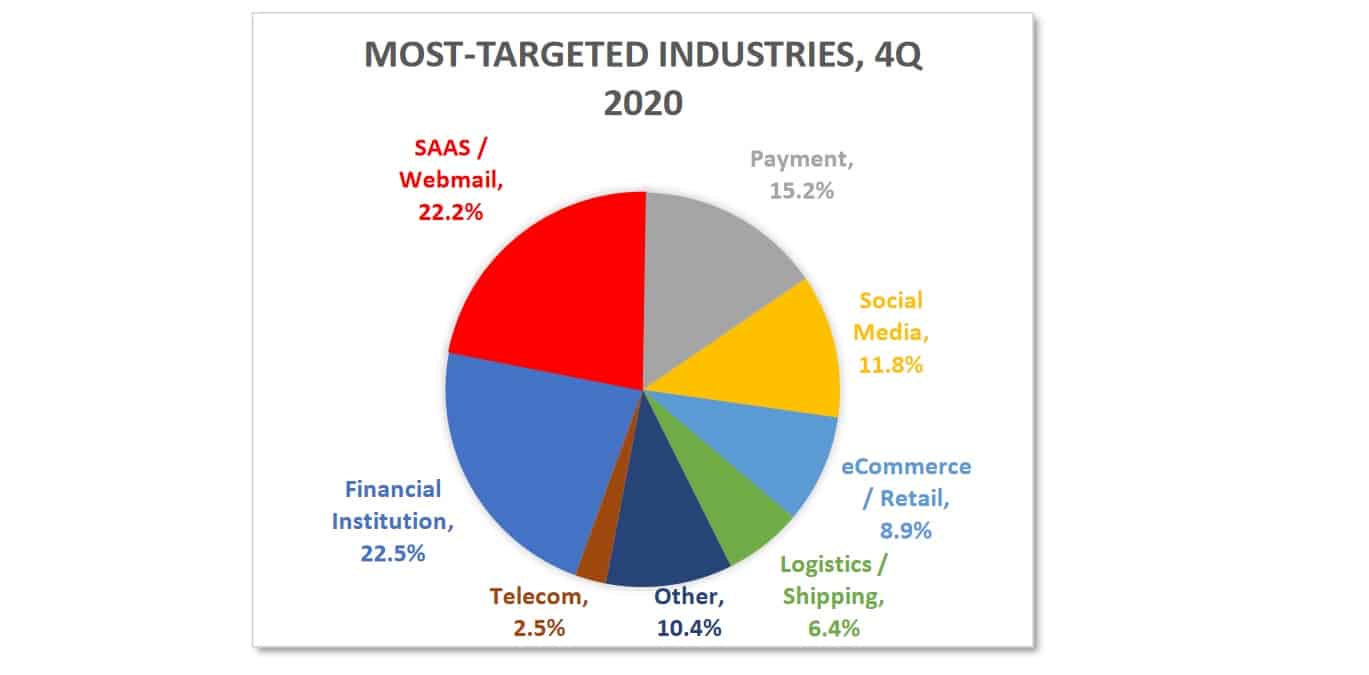} 
\caption{Most-Targeted Industries, 4Q,2020}
\end{figure}
\newline

The Figure 1 shows what industries are mainly targeted for phishing attacks. 
There’s been a marked change from previous years, though, with Software as a Service (Saas) and webmail attacks dropping from 31.4\% to 22.2\% in a single quarter. As such, financial institutions are now the most common target, accounting for 22.5\%. Meanwhile, attacks on eCommerce platforms and payment platforms have both risen by a few percent [4].\newline

The main objective of the paper is to develop a method that can be easily used by everyone to detect phishing websites accurately in real-time. The proposed technique extracts different features of a URL and predicts whether the site is safe or not. The method is then developed into a browser extension for testing, so that it is easy for the end-users to install and use. With the extension, the end-user will not able to click the websites directly which can save them from losing their personal information. They can hover on any link i.e., when a cursor is pointed to any link, the model shows the nature of the site i.e.,  safe site or not and then asks the user whether they want to visit the site. When the user approves then only the site is opened in a new window.
\newline 

The remaining paper is structured as follows: Literature Review is covered in Section II, followed by proposed methodology in section III. Performance analysis about the different classifier and evaluation results of testing method are covered in section IV. Finally, the Conclusion followed by future work is covered in section V.

\section{Literature Review}
Mohith et al [5] developed a special browser which is similar to the normal browsers but it has an additional module of detecting the phishing sites and warning the users about it. That module is named as "Intelligent Engine". This browser performs all the operations that a normal browser does. While surfing the website, the browser extracts the features of a URL and detects whether the site is safe or not. Here random forest is used for training the data and constructing a model which helps in prediction. Based on the training dataset, the model will be able to predict the new sites in little time. The Intelligent Engine detects the phishing URLs and renders the browsers showing a warning pop-up to the users.\newline \newline
One of the most popular methods is to manually detect phishing websites. As these types of attacks are now common these days, end-users must be trained with some basic knowledge about these attacks in order to avoid any loss. An model was proposed by Williams and Li [6] for evaluating the cognitive behavior of ACT-R. Based on the HTTP padlock's security indications, the authenticity of the web page is determined. Afroz and Greenstadt [7] designed a technology called "PhishZoo", in which the performance of the website before it loads is observed and then the profile of the website is also taken into the consideration while detecting the sites. In this method a list of vulnerable websites are stored and then they are compared with new websites. This is similar to  traditional techniques, whereas in this method comparison is done between the content of legitimate pages and malicious pages. \newline  \newline
Hu et al. [8] proposed a method for detecting phishing websites by analysing server log data. When a person visits a malicious website, the browser communicates with the legitimate website to collect resources. The legitimate server of the website registers the request in the registry, which is then used to identify the unauthorized request. Wu et al.[9] developed a system that integrates fuzzy logic with machine learning functions. They use information about the domain such as what are the subdomains, age of domains, expiry date of the certification and so on.\newline  \newline
Developers designed a software which is used to detect the phishing sites in real time and prevent them from accessing any kind of data. This software searches each and everything which is entering the system and if it detects any kind of danger, it blocks them and put in the black-list. The examples of such software are anti-virus and anti-malware. Usually in the browsers, malicious sites are blocked by listing techniques. In order to detect sites which have bypassed the listing techniques, Armano et al. [10] proposed a browser extension to detect the malicious sites. The extension gathers data from the website to identify the nature of the site, and then if the website is phishing, a warning message will appear on the screen. Similar to this method, Marchal et al.[11] proposed a plugin for the Firefox browser.\newline \newline
To solve the drawbacks of previous anti-phishing systems, Rao et al [12] proposed a classification model based on heuristic features collected from URLs, source code, and third-party services. The main objective is to extract most significant features from the URLs and give these features to machine learning classifiers for training and constructing a model. The features extracted are: URL Obfuscation features, Third-Party-based features, Hyperlink-based features. In the proposed method, they categorized the links into two types that is, local links and foreign links. The number of local links on any genuine site would exceed the number of foreign links. In this way detection of legitimate sites from multiple similar sites will be easy. Experimental results show that third-party service features have a considerable impact on a model's performance. Out of all oblique Random Forests, principal component analysis Random Forest (PCA-RF) achieved the hisghest accuracy of 99.55\%. \newline \newline
Jain et al [13] proposed a machine learning algorithm called logistic regression for detecting the websites by using the hyperlinks of the website. Usually phishing sites are short lived so we have to know more about the website in a limited time. The proposed method detects the links correctly and have an accuracy of 98.42\% which is quite effective when compared with the other algorithms. It doesn’t need third party services and works client-side. In this method hyperlinks are classified into twelve different categories such as total hyperlinks, no hyperlink, internal hyperlinks, external hyperlinks, internal error, external error, internal redirect, external redirect, null hyperlink, login form link, external/internal CSS, and external/internal favicon. Instead of using existing dataset, they developed a dataset. For that purpose, a web crawler is used to crawl through the website and extract all the hyperlinks present in the code, next they are grouped in the respective categories and then proposed algorithms work on them and the results generated will be binary codes. If the result is 1 then it is phishing else it is legitimate. This helps in detecting zero hour phishing attacks, and detecting websites written in any textual language. \newline \newline
Phishing attacks are of many types such as deceptive phishing,spear phishing, whale phishing ,URL phishing. kumar et al [14] discussed about URL phishing, usually phishing can be detected by blacklisting techniques but blacklists might not be exhaustive and they cannot detect any new attacks, so instead of blacklisting they used machine learning. URL phishing is changing names of the domain names or replacing certain letters with other letters which are used frequently by end users such as google can be written as goggle and this type of URLs cannot be detected by listing techniques. The basic parts of URL consists of protocol, domain name, sub-domain name, type of file and directory. As the end-users doesn't have any knowledge on these things, they can be deceived easily. The techniques used by attackers for URL phishing is cybersquatting and typosquatting which are also known as URL hijacking. For implementing the proposed method, they have taken a dataset and applied some machine learning algorithms to predict whether the URL is illegitimate or not. From the results Naïve-Bayes is the classifier which got high accuracy.

\section{Proposed Methodology}
Figure 2, shows the step-by-step procedure of the proposed phishing detection method. The first step in this method is extracting features from the URLs and giving them as input to six different machine learning classifiers for training.
\begin{figure}[htpb!]
\includegraphics[scale=0.4]{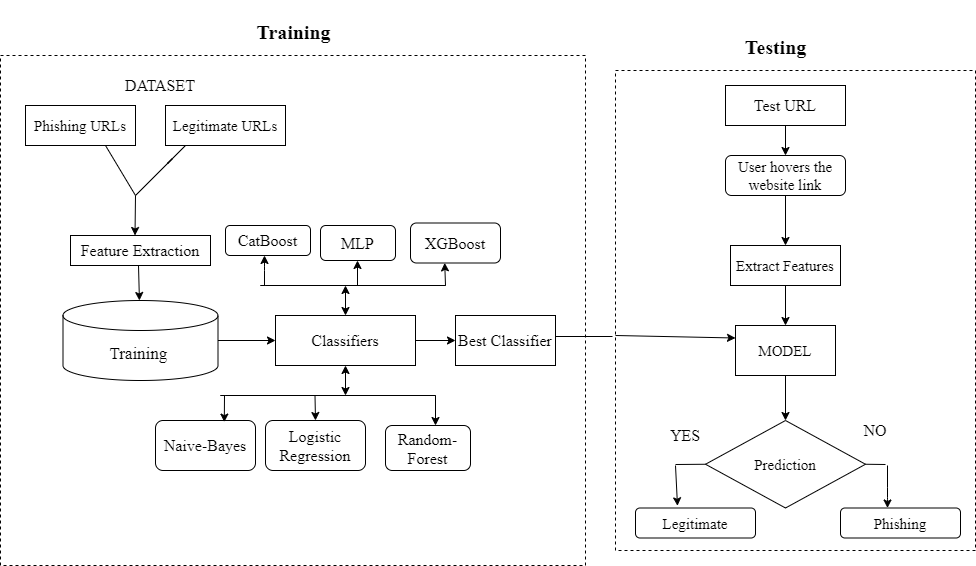} 
\caption{Flow-chart of Phishing detection method}
\end{figure}
After training, a model is developed by the best classifier, which helps in predicting the nature of the website. This method is developed into a browser extension for testing. The extension can be used by end-users while browsing. In this extension the user will not be able to click the website directly, they can just hover over the links. When the end-user points the cursor on any link then the model predicts the nature of the website and shows the result. In this way the end-user can minimize the loss of information. After knowing the authenticity of the website, it is up to the user to visit the website.

\subsection{Dataset Used}
We have used the following sources for collecting the data.
\begin{table}[!htpb]
\small
\centering
    \caption{Dataset used in our model} 
    \begin{tabular}{ m{2cm}  m{4cm}  m{2cm} }
\hline
Type &  Source & Sites\\
\hline
Legitimate & Majestic Million [15]  & 12669\\
\hline
Phishing  & Phishtank [16] & 11170\\
\hline
\end{tabular}
\end{table}
In Table 1 we can see the source of the data and the number of instances of the URLs taken to train the model.
\subsection{Feature Extraction}
There are different features of a website which will help in finding the authenticity of a website such as content-based [17], visual-based [18], URL-based[19] and so on. Here, we will use URL-based features to determine the novelty of a website. 
Along with the features, a label is added to each data. The label is used to distinguish between the classes. Here we have two classes i.e., whether the site is phishing site or not. It is a type of classification problem [21]. We will extract 23 different features of the URL which will help in developing the model. These features include:
\newline \newline
\setlength{\tabcolsep}{4em} 
\begin{tabular}{ l     l}
1. Have\_At       &        13.  Mouse\_Over  \\
2. URL\_Length     &        14. Right\_Click \\
3. UR\_Depth       &        15. Web\_Forwards \\
4. Redirection     &       16. having\_ip    \\
5. https\_Domain   &       17.  SSL           \\
6. TinyURL         &       18. https\_token  \\
7. Prefix/Suffix   &        19. sub\_domain   \\
8. DNS\_Record     &        20. request\_url  \\
9. Web\_Traffic    &        21. url\_anchor   \\
10. Domain\_Age    &        22. links         \\
11. Domain\_End    &        23. email         \\
12. iFrame         &        
\end{tabular}
\newline
The features are explained below:\newline
\textbf{1. Have\_At}: \newline
Using “@” symbol in the URL leads the browser to ignore everything preceding the “@” symbol and the real address often follows the “@” symbol. \newline
If the URL has '@' symbol, the value assigned to this feature is 1 (phishing) or else 0 (legitimate).\newline\newline
\textbf{2. URL\_Length}: \newline
Computes the length of the URL. Phishers can use long URL to hide the doubtful part in the address bar. In this project, if the length of the URL is greater than or equal 54 characters then the URL classified as phishing otherwise legitimate. \newline
If the length of URL $>=$ 54 , the value assigned to this feature is 1 (phishing) or else 0 (legitimate). \newline\newline
\textbf{3. URL\_Depth}:\newline
Computes the depth of the URL. This feature calculates the number of sub pages in the given url based on the '/'. The value of feature is a numerical based on the URL.\newline\newline
\textbf{4. Redirection}:\newline
Checks the presence of "//" in the URL. The existence of “//” within the URL path means that the user will be redirected to another website. Check for the position of "//".\newline
If the "//" is anywhere in the URL apart from after the protocol, thee value assigned to this feature is 1 (phishing) or else 0(legitimate).\newline\newline
\textbf{5. https\_Domain}:\newline
Checks for the presence of "http/https" in the domain part of the URL. The phishers may add the “HTTPS” token to the domain part of a URL in order to trick users. \newline
If the URL has "http/https" in the domain part, the value assigned to this feature is 1 (phishing) or else 0 (legitimate).\newline\newline
\textbf{6. TinyURL}:\newline
URL shortening is a method on the “World Wide Web” in which a URL may be made considerably smaller in length and still lead to the required web page.\newline
If the URL is using Shortening Services, the value assigned to this feature is 1 (phishing) or else 0 (legitimate)\newline\newline
\textbf{7. Prefix/Suffx}:\newline
Checking the presence of '-' in the domain part of URL. The dash symbol is rarely used in legitimate URLs.\newline
If the URL has '-' symbol in the domain part of the URL, the value assigned to this feature is 1 (phishing) or else 0 (legitimate).\newline\newline
\textbf{8. DNS\_Record}:\newline
For phishing websites, either the claimed identity is not recognized by the WHOIS database or no records founded for the hostname. \newline
If the DNS record is empty or not found then, the value assigned to this feature is 1 (phishing) or else 0 (legitimate).\newline\newline
\textbf{9. Web\_Traffic}:\newline
This feature measures the popularity of the website by determining the number of visitors and the number of pages they visit. If the domain has no traffic or is not recognized by the Alexa database, it is classified as “Phishing”.
If the rank of the domain $<$ 100000, the value of this feature is 1 (phishing) else 0 (legitimate).\newline\newline
\textbf{10. Domain\_Age}:\newline
This feature can be extracted from WHOIS database. Most phishing websites live for a short period of time. The minimum age of the legitimate domain is considered to be 12 months.\newline
If age of domain $<$ 12 months, the value of this feature is 1 (phishing) else 0 (legitimate).\newline\newline
\textbf{11. Domain\_End}:\newline
This feature can be extracted from WHOIS database. For this feature, the remaining domain time is calculated by finding the different between expiration time \& current time. The end period considered for the legitimate domain is 6 months or less. \newline
If end period of domain $<$ 6 months, the value of this feature is 1 (phishing) else 0 (legitimate).\newline\newline
\textbf{12. Iframe }:
IFrame is an HTML tag used to display an additional web page into one that is currently shown. Phishers can make use of the “iframe” tag and make it invisible i.e. without frame borders. \newline 
If the iframe is empty or response is not found then, the value assigned to this feature is 1 (phishing) or else 0 (legitimate).\newline\newline
\textbf{13. Mouse\_Over}:
Phishers may use JavaScript to show a fake URL in the status bar to users. To extract this feature, we must dig-out the web page source code, particularly the “onMouseOver” event, and check if it makes any changes on the status bar.\newline
If the response is empty or onmouseover is found then, the value assigned to this feature is 1 (phishing) or else 0 (legitimate).
\newline\newline
\textbf{14. Right\_Click}:
Phishers use JavaScript to disable the right-click function, so that users cannot view and save the web page source code. This feature is treated exactly as “Using onMouseOver to hide the Link”. Nonetheless, for this feature, we will search for event “event.button==2” in the web page source code and check if the right click is disabled.\newline
If the response is empty or onmouseover is not found then, the value assigned to this feature is 1 (phishing) or else 0 (legitimate). \newline\newline
\textbf{15. Web\_Forwards}:
The fine line that distinguishes phishing websites from legitimate ones is how many times a website has been redirected. In our dataset, we find that legitimate websites have been redirected one time max. On the other hand, phishing websites containing this feature have been redirected at least 4 times.
If the response is empty or if the url has been redirected more than 3 times then this feature is 1 (phishing) or else 0 (legitimate).\newline\newline
\textbf{16. having\_ip}:
If an IP address is used as an alternative of the domain name in the URL, such as “http://115.102.3.123/home.html”, users can be sure that someone is trying to steal their personal information.\newline
If the url has ip address in the domain part then the feature is 1 (phishing) or
else 0 (legitimate). \newline\newline
\textbf{17. SSL}:
The existence of HTTPS is very important in giving the impression of website legitimacy, but this is clearly not enough. Checking the certificate assigned with HTTPS including the extent of the trust certificate issuer, and the certificate age.\newline
If "https" is used and Issuer Is Trusted \& and Age of Certificate $>=$ 1 Years then the feature is 0 (Phishing) else if Using https and Issuer Is Not Trusted then the feature is -1 (suspicious) or 1 (Phishing).\newline\newline
\textbf{18. https\_token}:
Checks for the presence of "http/https" in the sub-domain part of the URL.\newline
If the URL has "http/https" in the domain part, the value assigned to this feature is 1 (phishing) or else 0 (legitimate).\newline\newline
\textbf{19. sub\_domain}:
To produce a rule for extracting this feature, we firstly have to omit the (www.) from the URL which is in fact a sub domain in itself. Finally, we count the remaining dots. If the number of dots is greater than one, then the URL is classified as “Suspicious” since it has one sub domain. However, if the dots are greater than two, it is classified as “Phishing” since it will have multiple sub domains. Otherwise, if the URL has no sub domains, we will assign “Legitimate” to the feature.\newline
If Dots In Domain Part=1 then feature is 0 (Legitimate) else Dots In Domain Part=2 then feature is -1 (Suspicious) or then it is 1 (Phishing).\newline\newline
\textbf{20. request\_url}:
Request URL examines whether the external objects contained within a web page such as images, videos and sounds are loaded from another domain. In legitimate web pages, the web page address and most of objects embedded within the web page are sharing the same domain.\newline.
If the \% of Request URL $<$22\% then the feature is 1 (Legitimate) else if \%of Request URL$>=22$\% and 61\%→ Suspicious,Otherwise feature is 1 (Phishing).\newline\newline
\textbf{21. url\_anchor}:
<a> (anchor tag) in html code refers to which link it is redirecting or pointing. If there are such <a> tags whose address is empty or whose address is different from the main domain then it is considered as malicious.\newline
If \% of URL Of Anchor $<$31\% then it is 0 (Legitimate) else if \% of URL Of Anchor $>=$31\% And $<=$67\% then it is -1 (Suspicious), Otherwise it is 1 (Phishing).\newline\newline
\textbf{22. links}:
It is common for legitimate websites to use <Meta> tags to offer metadata about the HTML document; <Script> tags to create a client side script; and <Link> tags to retrieve other web resources. It is expected that these tags are linked to the same domain of the web page.\newline
If the \% of Links in "<Meta>","<Script>" and "<"Link>""$<$17\% then the feature is 1 (Legitimate) else \% of Links in <Meta>","<Script>" and "<"Link>" $>=$17\% And $<=$ 81\% then the feature is -1 (Suspicious), Otherwise→ it is 1 (Phishing).\newline\newline
\textbf{23. email}:
Web form allows a user to submit his personal information that is directed to a server for processing. A phisher might redirect the user’s information to his personal email. To that end, a server-side script language might be used such as “mail()” function in PHP. One more client-side function that might be used for this purpose is the “mailto:” function.\newline.
If there is mail() or mailto: then the feature is 1 (phishing) else 0 (legitimate).\newline\newline

\section{Performance Analysis}
\subsection{Training the Model \& Comparison of Results}
The extracted features are given as input to different machine learning classifiers for training. After training, we compare the results of the classifiers and then choose the appropriate one. The chosen classifier is used to develop a machine learning model which will be used for evaluating the nature of the websites. To create a more effective model, we have used GridSearchCV along with ten-fold cross-validation, so that the entire data is trained and tested equally [20]. 
The different machine learning classifiers used are  Random-Forest, CatBoost, XGBoost, Multilayer perceptron and so on. \newline
Table 2 shows the performance comparison of the classifiers along with the results. Performance of the classifiers is evaluated in terms of accuracy, precision, recall and F1-score.
\setlength{\tabcolsep}{0em}
\begin{center}
\begin{table*}[!htpb]
      \caption{Performance comparison of different Classifiers}
\begin{tabular}{| C{3em} | C{5cm}| C{2cm} | C{2cm} | C{2cm} | C{2cm} |} 
\hline
S.No & Classifier & Accuracy  & Precision & Recall  & F1-score \\ 
\hline
1 & Naive-Bayes &  0.7235 & 0.80  &   0.72  &    0.69  \\ 
\hline
2 & Logistic Regression & 0.8919 & 0.89 &  0.89 &  0.89    \\
\hline
3 & Multilayer perceptron(MLP) &  0.9351 & 0.94    &  0.94     & 0.94 \\
\hline
4 & XGBoost & 0.9377 &  0.94  &    0.94     & 0.94 \\
\hline
5 & CatBoost &  0.9387 &   0.94  &    0.94     & 0.94 \\ 
\hline
6 & Random-forest  & 0.9501 & 0.95 & 0.95 & 0.95 \\
\hline
\end{tabular}
\end{table*}
\end{center}
Among all the classifiers, Random-Forest achieved the highest accuracy i.e, 95\%, so this is used for developing the model which will help in prediction.

\subsection{Results and Discussion}
The above method is developed into a plugin and tested in the chrome extension. When the user wants to visit any website the user have to click the link and that link redirects to the website. In the proposed method the clicked link is taken as input and stored in a variable. Then the variable is sent to the python server locally where the machine learning model predicts the nature of the website based on its features. Figure 3, shows the test URL in the form of pop-up box. After the prediction of the URL, the result is shown. Figure 4, shows a pop-up box indicating the URL is safe, and Figure 5, shows that the URL tested is a deceptive site. After the nature of the site is shown then a confirm-box appears. Figure 6 shows a confirm-box asking the user whether they want to visit the site or not. If the user wants to visit the site, then it is opened in a new window. A confirm-box appears after the predicted result. The extension is linked to the local database to store the hyperlinks along with the predicted result. This data can be used for increasing the performance of the method.
\begin{figure}[htpb!]
  \centering 
\includegraphics[scale=0.4]{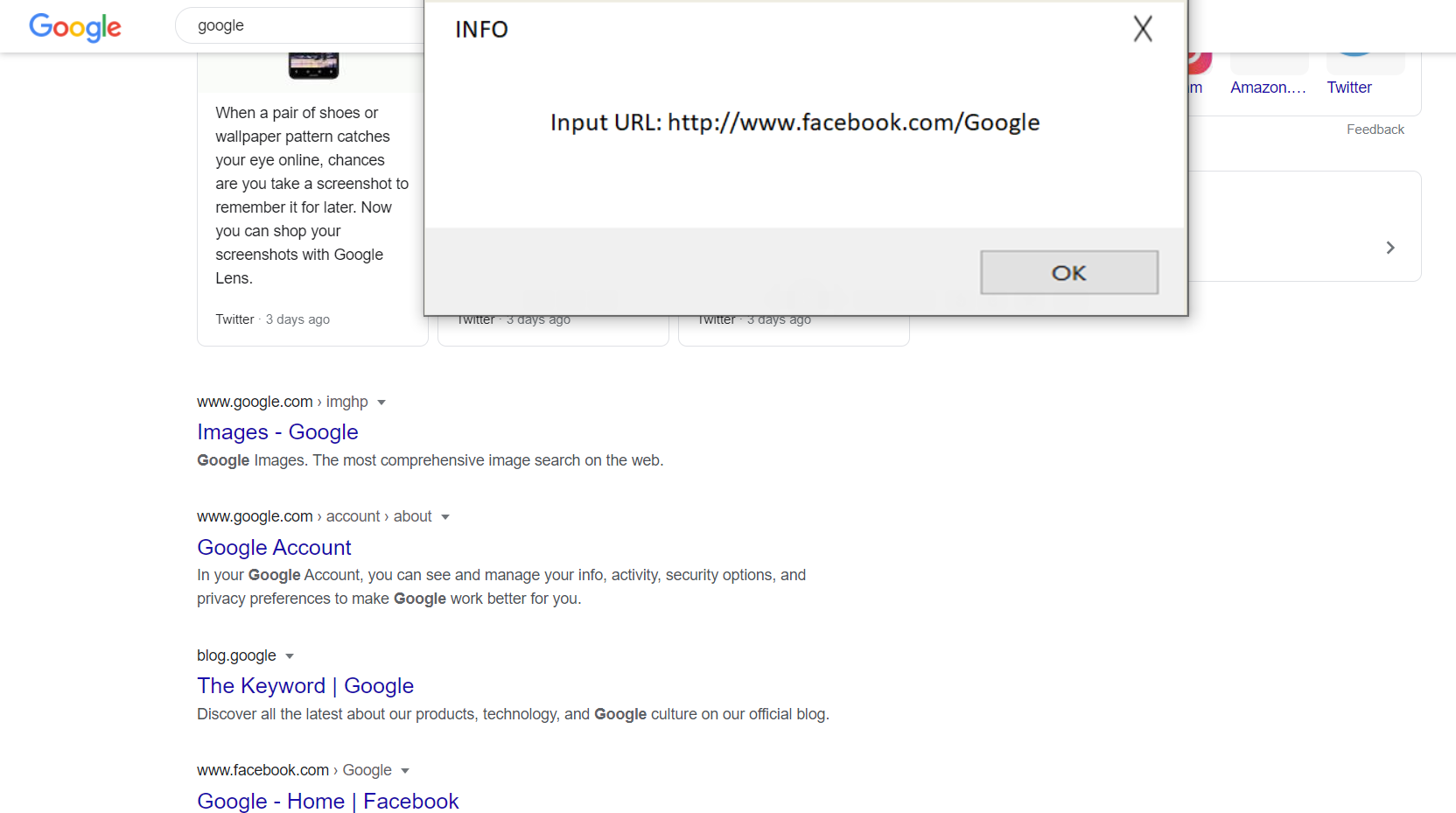} 
\caption{Pop-up of the Test URL}
\end{figure}

\begin{figure}[htpb!]
\includegraphics[scale=0.7]{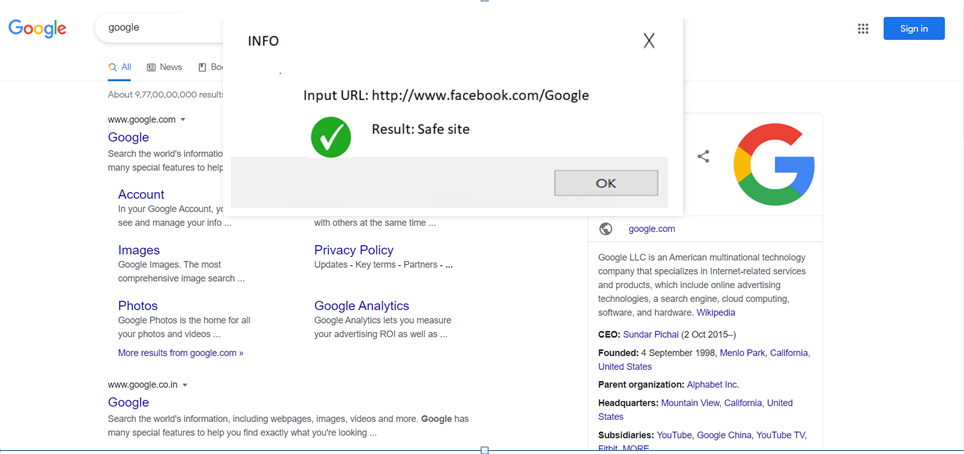} 
\caption{Pop-up showing that URL is safe}
\end{figure}

\begin{figure}[htpb!]
  \centering 
\includegraphics[scale=0.8]{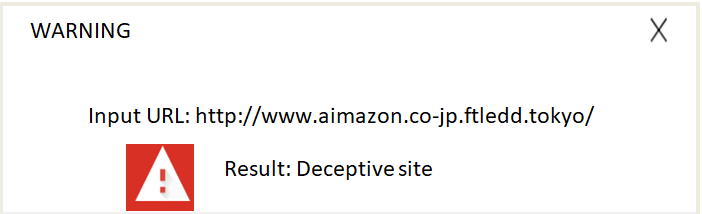} 
\caption{Pop-up showing that URL is deceptive}
\end{figure}

\begin{figure}[htpb!]
  \centering 
\includegraphics[scale=0.8]{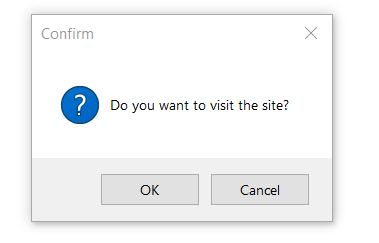} 
\caption{Confirm-box asking whether the user wants to visit the site}
\end{figure}


\newpage
\section{Conclusion}
In this paper we proposed a method where the authenticity of the website is known without visiting them. Then this method is developed into a browser extension so that end-users can use it while surfing the net. In this the user just has to point the cursor over any link in any site to know the nature of the site. After knowing the nature of the site, if the user wants to visit the site then the site will be opened otherwise it will not. For constructing the machine learning model we have used six different classifiers such as Naive-Bayes, Logistic Regression, Random Forest, CatBoost, XGBoost and Multilayer perceptron. After comparing with the other classifiers, the Random forest is selected as it gives the highest accuracy i.e., 95\%. For the continuous improvement of the model, the extension is linked with the database to store the URL along with the predicted result. As the data is being stored, we can find incorrect results from the database and give the new data to the dataset to increase the performance of the model.\newline
In future, we plan to work on developing a technique which can detect the phishing in deep and dark websites. 
 \newpage

\section*{References}
[1] https://apwg.org/ \newline \newline
[2]Heartfield, Ryan \& Loukas, George. (2016). A Taxonomy of Attacks and a Survey of Defence Mechanisms for Semantic Social Engineering Attacks. ACM Computing Surveys. 48. 10.1145/2835375. \newline \newline
[3]https://www.imperva.com/learn/application-security/phishing-attack-scam/\newline \newline
[4]https://www.comparitech.com/blog/vpn-privacy/phishing-statistics-facts/\newline\newline
[5]HR, M., MV, A., S, G. et al. Development of anti-phishing browser based on random forest and rule of extraction framework. Cybersecur 3, 20 (2020). https://doi.org/10.1186/s42400-020-00059-1 \newline \newline
[6]Williams N, Li S (2017) Simulating human detection of phishing websites: an investigation into the applicability of the ACT-R cognitive behaviour architecture model. In: 2017 3rd IEEE international conference on cybernetics (CYBCONF). https://doi.org/10.1109/cybconf.2017.7985810\newline \newline
[7]Afroz S, Greenstadt R (2011) PhishZoo: detecting phishing websites by looking at them. In: 2011 IEEE fifth international conference on semantic computing, Palo Alto, CA, pp 368–375. https://doi.org/10.1109/ICSC.2011.52. \newline \newline
[8]Hu J, Zhang X, Ji Y, Yan H, Ding L, Li J, Meng H (2016) Detecting phishing websites based on the study of the financial industry webserver logs. In: 2016 3rd international conference on information science and control engineering (ICISCE), pp 325–328. https://doi.org/10.1109/icisce.2016.79 \newline \newline
[9]Wu C, Kuo C, Yang C (2019) A phishing detection system based on machine learning. In: 2019 international conference on intelligent computing and its emerging applications (ICEA), Tainan, Taiwan, pp 28–32. https://doi.org/10.1109/ICEA.2019.8858325 \newline \newline
[10]Armano G, Marchal S, Asokan N (2016) Real-time client-side phishing prevention add-on. In: 2016 IEEE 36th international conference on distributed computing systems (ICDCS), pp 777–778. https://doi.org/10.1109/icdcs.2016.44 \newline \newline
[11]Marchal S, Armano G, Grondahl T, Saari K, Singh N, Asokan N (2017) Off-the-hook: an efficient and usable client-side phishing prevention application. IEEE Trans Comput 66(10):1717–1733. https://doi.org/10.1109/tc.2017.2703808\newline \newline
[12]Rao, R. S. and Pais, A. R. (2019). Detection of phishing websites using an efficient feature-based machine learning framework. Neural Computing and Applications (Springer), 1-23. [DOI: https://doi.org/10.1007/s00521- 017-3305-0]\newline 
[13]Jain, A. K., \& Gupta, B. B. (2018). “A machine learning based approach for phishing detection using hyperlinks information”. Journal of Ambient Intelligence and Humanized Computing. doi:10.1007/s12652-018-0798-z \newline 
[14]Kumar, J., Santhanavijayan, A., Janet, B., Rajendran, B., \& BS, B. (2020). Phishing Website Classification and Detection Using Machine Learning. 2020 International Conference on Computer Communication and Informatics (ICCCI). doi:10.1109/iccci48352.2020.9104161\newline
[15]https://majestic.com/reports/majestic-million\newline \newline
[16]https://phishtank.com/developer\_info.php \newline \newline
[17]G. J. W. Kathrine, P. M. Praise, A. A. Rose and E. C. Kalaivani, "Variants of phishing at-tacks and their detection techniques," 2019 3rd International Conference on Trends in Electronicsand Informatics (ICOEI), Tirunelveli, India, 2019, pp. 255-259, doi: 10.1109/ICOEI.2019.8862697. \newline \newline
[18]Sahar Abdelnabi, Katharina Krombholz, and Mario Fritz. 2020. VisualPhishNet: Zero-DayPhishing Website Detection by Visual Similarity. In Proceedings of the 2020 ACM SIGSAC Con-ference on Computer and Communications Security (CCS ’20). Association for Computing Ma-chinery, New York, NY, USA, 1681–1698. DOI:https://doi.org/10.1145/3372297.3417233\newline \newline
[19]M. Korkmaz, O. K. Sahingoz and B. Diri, "Feature Selections for the Classification of Web-pages to Detect Phishing Attacks: A Survey," 2020 International Congress on Human-ComputerInteraction, Optimization and Robotic Applications (HORA), Ankara, Turkey, 2020, pp. 1-9, doi:10.1109/HORA49412.2020.9152934. \newline \newline
[20]https://stackabuse.com/cross-validation-and-grid-search-for-model-selection-in-python/ \newline \newline
[21]https://www.edureka.co/blog/classification-in-machine-learning/ \newline \newline

\end{document}